\documentclass[pra,aps,twocolumn,showpacs,superscriptaddress,amsfonts,amsmath,floatfix]{revtex4-2}

\usepackage{graphicx}
\usepackage{amsmath}
\usepackage{amsfonts}
\usepackage{amssymb}
\usepackage{tikz}
\usepackage[font=small, justification=RaggedRight, singlelinecheck=off]{caption}
\usepackage{subcaption}
\usetikzlibrary{decorations.pathmorphing,patterns}

\begin{document}

\title{An exactly solvable model for anyons with non-Abelian flux}

\author{Bruno Klajn}
\email{bk@phy.hr}
\affiliation{Department of Physics, Faculty of Science, University of Zagreb, Bijeni\v{c}ka c. 32, 10000 Zagreb, Croatia}

\author{Silvije Domazet}
\affiliation{Department of Physics, Faculty of Science, University of Zagreb, Bijeni\v{c}ka c. 32, 10000 Zagreb, Croatia}

\author{Dario Juki\'{c}}
\affiliation{Faculty of Civil Engineering, University of Zagreb, A. Ka\v{c}i\'{c}a  Mio\v{s}i\'{c}a 26, 10000 Zagreb, Croatia}

\author{Hrvoje Buljan}
\affiliation{Department of Physics, Faculty of Science, University of Zagreb, Bijeni\v{c}ka c. 32, 10000 Zagreb, Croatia}

\date{\today}

\begin{abstract}
We present an exactly solvable model for synthetic anyons carrying non-Abelian flux.  
The model corresponds to a two-dimensional electron gas in a magnetic field with a specific spin interaction term, 
which allows only fully aligned spin states in the ground state; the ground state subspace is thus two-fold degenerate. 
This system is perturbed with identical solenoids carrying a non-Abelian gauge potential. 
We explore dynamics of the ground state as these solenoids are adiabatically braided and 
show they behave as anyons with a non-Abelian flux. 
Such a system represents a middle ground between the ordinary Abelian anyons and the fully non-Abelian anyons.
\end{abstract}

\maketitle

\section{Introduction}

One of the peculiarities of two-dimensional (2D) quantum systems is the existence of particles which are 
neither bosons nor fermions, but whose exchange statistics interpolates between the two. 
Since these particles can have any statistical phase, they were termed anyons~\cite{Wilczek1982}. 
Anyons are a direct consequence of the fact that the process of exchanging two identical particles is not merely a permutation of their quantum numbers, but an adiabatic interchange of the positions of the particles. 
The dimension of a quantum system determines the possible 
indistinguishable particles it can support~\cite{Laidlaw1971, Leinaas1977}. 
The topology of 2D systems, unlike the more familiar 3D systems, makes the interchange 
of two indistinguishable quantum particles nontrivial, leading to the very existence of anyons. 
For two identical noninteracting anyons whose energy spectrum is nondegenerate, the exchange 
phase is a special case of the geometrical (Berry) phase~\cite{Berry1984}. 
In this case, the Berry phase is not only geometrical, but is also topological in nature. These type of anyons are called Abelian anyons. 
On the other hand, a prerequisite for non-Abelian anyons is a degenerate ground state manifold~\cite{Nayak2008}, 
such that braiding of anyons corresponds to a unitary transformation (rotation) within the subspace manifold. 
This rotation is described by a unitary Wilczek-Zee matrix~\cite{WZ1984}, which is a direct generalization of the Berry phase.

Although anyons do not exist (or at least they were not found) as elementary particles, 
they have been experimentally realized as emergent quasiparticle excitations in condensed matter systems. 
The most notable phenomenon in which anyons appear as emergent quasiparticles is the 
Fractional Quantum Hall Effect (FQHE)~\cite{Tsui1982, Laughlin1983, Arovas1984, Camino2005, Feldman2021}. 
Anyons are found, as well, in Kitaev spin systems~\cite{Kitaev2003, Kitaev2006, Dai2017, Klanjsek2018} and 
Majorana zero modes~\cite{DasSarma2015, Mourik2012}.
Both the FQHE and the Kitaev model, together with other approaches including synthetic gauge potentials, 
have been proposed as routes for experimental realizations of anyons in ultracold atomic 
gases~\cite{Paredes2001, Zhang2014, Duan2003, Jiang2008, Burrello2010, Andrade2021, Baldelli2021}. 
Alternative schemes have been proposed to achieve FQH states of light~\cite{Kapit2014, Umucalilar2017}. 
Anyonic statistics has been simulated in photonic quantum simulators~\cite{Lu2009, Pachos2009}, superconducting quantum circuits~\cite{Zhong2016}, and with the use of nuclear magnetic resonance~\cite{Li2017}. Scanning tunneling microscopy (STM) has been proposed for imaging anyons~\cite{Papic2018}, and it plays a key role in the experimental search for Majorana zero modes~\cite{Jaeck2021}. Novel ideas have also been put forward~\cite{Brooks2021} according to which anyons could be experimentally realized on a non-trivial 2D surface (a sphere), using molecular impurities.
Recently, successful experiments were carried out in which FQH states were observed to be anyonic using interferometry \cite{Nakamura2020} and in a particle collision in the beamsplitter \cite{Bartolomei2020}. Also, direct braiding of anyons was achieved using photonic waveguides \cite{Noh2020}.

The most attractive property of (non-Abelian) anyons is their topological stability which makes them a highly desirable platform for fault tolerant quantum computing~\cite{Kitaev2003, Nayak2008}. 
However, an experimental implementation of this idea still evades effective realization~\cite{Nayak2008, DasSarma2015, Barreiro2011}. 
Recent progress in synthesizing and observing non-Abelian gauge fields~\cite{Yang2019} invites further research in this and related directions. 
The conventional scheme of anyons appearing only as a result of an interaction between constituents of a system has been challenged and other ideas have also been put forward, in which anyons would appear as a result of coupling noninteracting (or weakly interacting) electrons to either a topologically nontrivial background, or a topologically nontrivial external perturbations. In this regard, anyons were proposed to be realized in a system of an artificially structured type-II superconducting film~\cite{Weeks2007, Rosenberg2009} adjacent to a 2D electron gas (2DEG) in the Integer Quantum Hall Effect (IQHE)~\cite{Klitzing1980, Laughlin1981}, in IQHE magnets~\cite{Rahmani2013}, 
in topological defects in graphene~\cite{SeradjehPRL2008}, and by sandwiching a charged magnetic dipole between 
two semi-infinite blocks of a high permeability metamaterial~\cite{Todoric2018}. 
Recently, a theoretical model for synthetic Abelian anyons in a noninteracting system was proposed along these lines~\cite{Lunic2020}, in which specially tailored localized probes are brought into the IQHE setup and shown to have anyonic properties. 
It should be pointed out that these synthetic Abelian anyons are not emergent quasiparticles (see Refs.~\cite{Lunic2020, Todoric2020}
and the commentary on Refs.~\cite{Weeks2007, Rosenberg2009} therein)

In this paper we follow the aforementioned model~\cite{Lunic2020} and generalize it in two aspects. First, we introduce the spin interaction between the electrons in such a way to produce a degenerate ground state, the arena for non-Abelian physics, and second, we equip the probes with a non-Abelian gauge potential. For the model to remain exactly solvable, it is necessary that the non-Abelian features of the gauge potential are as simple as possible. The price we pay for this simplicity is that the anyons that form are not fully non-Abelian, but merely carry the non-Abelian flux. However, this is an important step in reaching the goal of true non-Abelian anyons.

\section{The system under consideration}

\begin{figure}%
\includegraphics{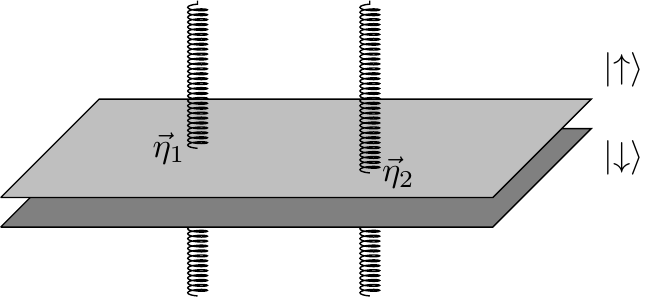}
\caption{A sketch of the system consisting of two solenoids with non-Abelian flux immersed in the planar arrangement of electrons with two internal degrees of freedom.}%
\label{figura1}
\end{figure}

We consider a system consisting of $N_\text{e}$ identical charged particles of mass $\mu$ and charge $q$. 
Each particle has two internal degrees of freedom. For clarity of the presentation, we will call the particles electrons, 
and the internal degree of freedom spin, having in mind that both the ''charge'' and "spin'' of the ''electron'' 
could be of synthetic origin in the sense that the pertinent Hamiltonian could be experimentally realized 
on a different platform (e.g., ultracold atomic gases). 
The electrons are confined to move in the $xy$ plane with the uniform magnetic field $\vec{B}_0 = B_0 \hat{z}$, 
with $B_0 >0$, normal to the plane. We will use the vector potential in a symmetric Coulomb gauge to describe this 
magnetic field, $\vec{A}_0 = \frac{1}{2}\vec{B}_0 \times \vec{r}$, where $\vec{r}$ is a 2D position vector taken 
from some origin in the plane. 
In addition to the uniform magnetic field, the plane is pierced by $N$ identical thin solenoids located at 
positions $\vec{\eta}_k$, as sketched in Fig.~\ref{figura1}. The standard Abelian solenoid located at 
$\vec{\eta}_k$ produces a vector potential
\begin{equation*}
\vec{A}'_k = \frac{\Phi}{2 \pi} \frac{\hat{z} \times (\vec{r} - \vec{\eta}_k)}{(\vec{r} - \vec{\eta}_k)^2},
\label{SolPot}
\end{equation*}
where $\Phi$ is the magnetic flux through the solenoid. 
In comparison to the system studied in \cite{Lunic2020}, the novel ingredient here
is that the vector potential of the solenoids is non-Abelian. 
We write it in the form
\begin{equation}
\vec{\mathcal{A}}_k = - \frac{\hbar}{q} \frac{\hat{z} \times (\vec{r} - \vec{\eta}_k)}{(\vec{r} - \vec{\eta}_k)^2} \mathcal{M} \equiv \vec{A}_k \mathcal{M},
\label{nonAbPot}
\end{equation}
where $\mathcal{M}$ is a constant, dimensionless $2 \times 2$ Hermitian matrix operating on the spin degrees of freedom. The potential having this form was first introduced in \cite{Wu1975}, and a physical realization of it has recently been discussed in \cite{Zygelman2021}. The matrix $\mathcal{M}$ can be readily diagonalized in the form
\begin{equation}
\mathcal{M} = 
\begin{pmatrix}
\alpha & 0 \\
0 & \beta
\end{pmatrix},
\label{DiagM}
\end{equation}
with real entries $\alpha$ and $\beta$. The matrix $\mathcal{M}$ is a linear combination of the unit matrix $\mathcal{I}$ 
and the third Pauli matrix $\sigma_3$, $\mathcal{M} = \mathcal{I} (\alpha + \beta)/2 + \sigma_3 (\alpha - \beta)/2$. 
In this basis, the spin-up electron sees the solenoid carrying magnetic flux proportional to $\alpha$, 
while the spin-down electron similarly sees the magnetic flux of the solenoid to be proportional to $\beta$. 
Therefore, different spin orientations couple differently to the solenoids. 
However, we keep the external magnetic field $\vec{B}_0$ Abelian, so that both spin orientations couple to it the same way. 
Having said that, we are in position to write a single particle Hamiltonian of the $j$-th electron in the form
\begin{equation}
H_{1,j} = \frac{1}{2\mu} \left\{ \mathcal{I} \left[\vec{p}_j - q \vec{A}_0 (\vec{r}_j) \right] - \mathcal{M} \sum_{k=1}^N q \vec{A}_k (\vec{r}_j) \right\}^2.
\label{spHam}
\end{equation}
The total Hamiltonian $H = H_0 + V$ consists of $H_0$, describing $N_\text{e}$ noninteracting electrons in an external non-Abelian vector potential, which is just the sum of single particle Hamiltonians
\begin{equation}
H_0 = \sum_{j=1}^{N_\text{e}} H_{1,j} ,
\label{nonint}
\end{equation}
and $V$ describes the spin interaction between electrons. 

Here we assume that $V$ acts on the spin states as follows. 
For the state with all spins up, 
$V\left| \uparrow \right\rangle =\epsilon_0 \left| \uparrow \right\rangle$, and 
equivalently for the state with all spins down, 
$V\left| \downarrow \right\rangle =\epsilon_0 \left| \downarrow \right\rangle$; 
we take $\epsilon_0=0$ without losing any generality. 
At the same time, any state with two or more unaligned spins is much higher in energy. 
Therefore, all unaligned spin states are excluded from further analysis.


This form of the interaction $V$ assures that there is a gap between the two spin states, $\left| \uparrow \right\rangle$ and $\left| \downarrow \right\rangle$, and the other $2^{N_\text{e} - 1}$ spin states. 
In simple words, the spin interaction term $V$ selects only two spin states from the Hilbert space to form a twofold degenerate ground state manifold, which provides a possibility for the non-Abelian dynamics.
With this in mind, we may write $V$ in the form
\begin{equation}
V = \Delta \left( \mathbb{I} - \left| \uparrow \right\rangle \left\langle \uparrow \right| -  \left| \downarrow \right\rangle \left\langle \downarrow \right| \right),
\end{equation}
where $\mathbb{I}$ is the identity operator and $\Delta > 0$ is an energy defect, much greater than the ground state energy.
We neglect the Zeeman splitting between the $\left| \uparrow \right\rangle$ and $\left| \downarrow \right\rangle$ states. 
Finally, we do not consider the effect of Coulomb repulsion between the electrons. This formally means that we are considering the limit $q \to 0$, with $\alpha$ and $\beta$, as well as the magnetic length $l_B = \sqrt{\hbar/|q| B_0}$ and cyclotron frequency $\omega_B = |q| B_0 / \mu$ held constant. 
While this model may be difficult to experimentally realize in a realistic setting, we proceed with the analysis 
as it will provide a useful information on the possibility to obtain synthetic non-Abelian anyons 
following the proposed scheme.

\section{Ground state dynamics}

Having constructed a Hamiltonian such that (at most) two of its spin states, $\left| \uparrow \right\rangle$ and $\left| \downarrow \right\rangle$, contribute to the ground state, we proceed to calculate that very state. As we have just argued, for ground state (GS) considerations, one can effectively disregard the potential $V$ and use the Hamiltonian $H_\text{GS} = H_0$. Writing the GS ket in the form
\begin{equation}
| \psi \rangle_{\text{GS}} = \psi_1 \left| \uparrow \right\rangle + \psi_2 \left| \downarrow \right\rangle,
\label{gsKet}
\end{equation}
where the wavefunctions $\psi_{1,2} (\{\vec{r}_i\}, \{\vec{\eta}_j\})$ depend on the positions of all the electrons and solenoids, 
the Schr\"odinger equation (SE) $H_0 | \psi \rangle_{\text{GS}} = E_0 | \psi \rangle_{\text{GS}}$ reduces to a set of two scalar equations
\begin{align}
\frac{1}{2\mu} \sum_{j=1}^{N_\text{e}}\left\{\left[\vec{p}_j - q \vec{A}_0 (\vec{r}_j) \right] - \alpha \sum_{k=1}^N q \vec{A}_k (\vec{r}_j) \right\}^2 \psi_1 = E_0 \psi_1, 
\label{scaleq1} \\
\frac{1}{2\mu} \sum_{j=1}^{N_\text{e}}\left\{\left[\vec{p}_j - q \vec{A}_0 (\vec{r}_j) \right] - \beta \sum_{k=1}^N q \vec{A}_k (\vec{r}_j) \right\}^2 \psi_2 = E_0 \psi_2,
\label{scaleq2}
\end{align}
which are just the SE for the system under consideration with the Abelian vector potential. Therefore, the ground state equations for our non-Abelian model simplifies to two separate Abelian problems, each of which couples to the same solenoids with different strength. 
This problem was studied in Ref.~\cite{Lunic2020} (see the Appendix of that paper for details of calculation), we only briefly outline the key steps leading to the solution. 

First of all, let us consider a single particle ($N_\text{e} = 1$) solution of Eq.~\eqref{scaleq1} in the case of a single Abelian solenoid ($N=1$). This is a typical IQHE setup with a single perturbation. It is no surprise that the energy spectrum of Eq.~\eqref{scaleq1} is split into Landau levels, with additional solenoid-induced states appearing in mid-gaps. Since the transformation $\alpha \to \alpha + 1$ corresponds to the gauge transformation of the vector potential, we assume that $\alpha \in \langle 0,1\rangle$. In that case, the lowest Landau level (LLL) single particle state is of the form
\begin{equation}
\psi_{LLL}^{1}(z) = \frac{\overline{z-\eta}\hphantom{^\alpha}}{|z-\eta|^\alpha} \bar{z}^m \exp(-|z|^2/4l_B^2),
\label{LLL}
\end{equation}
where we have switched to complex notation, so that, instead of $\vec{r}_j = x_j \hat{x} + y_j \hat{y}$ and $\vec{\eta}_k = \eta_{k,x} \hat{x} + \eta_{k,y} \hat{y}$, we write $z_j = x_j + i y_j$ and $\eta_k = \eta_{k,x} + i \eta_{k,y}$, respectively. Here, $m$ is a non-negative integer, labeling the states within the LLL. These states are degenerate, with energies $E_{LLL} = \hbar \omega_B/2$, independent of the solenoid strength $\alpha$. This fact is especially important, because it ensures that the ground state energies of systems in Eqs.~\eqref{scaleq1} and \eqref{scaleq2} will be the same, regardless of the fact that the systems differ in the couplings $\alpha$ and $\beta$. It should be noted that the energies of excited states in general depend on $\alpha$ (or $\beta$), which makes simultaneously solving Eqs.~\eqref{scaleq1} and \eqref{scaleq2} much more difficult. However, for our purposes, it is sufficient to consider the simpler, ground state case.

After explicitly solving the SE in the case of a single solenoid, we move on to the case of multiple solenoids. Now the complexity of the SE is such that we have to resort to some educated guesswork to obtain the ground state solution. From the form of Eq.~\eqref{LLL}, it seems that, at least for $N=1$, the effect of the solenoid is to multiply the LLL for the IQHE with the factor $(\overline{z-\eta})/|z-\eta|^\alpha$, and this also turns out to work for multiple solenoids, as the ansatz
\begin{equation}
\psi_{LLL}^{N}(z) = \prod_{k=1}^N \left(\frac{\overline{z-\eta_k}\hphantom{^\alpha}}{|z-\eta_k|^\alpha}\right) \bar{z}^m \exp(-|z|^2/4l_B^2)
\label{ansatz}
\end{equation}
satisfies the single particle SE for $N$ solenoids located at $\eta_k$. The ground state energy is unchanged by the introduction of multiple solenoids, while the excited states remain unobtainable by this method.

Having solved the single particle problem for arbitrary number of solenoids, we can construct the many-body wavefunction by forming a Slater determinant with the solutions in \eqref{ansatz}. We assume that there are only $N_\text{e}$ states in the LLL and that electrons fill all of them. The determinant obtained in this manner is of Vandermonde form and can be explicitly calculated. With this, we arrive at the solution for the ground state of Eq.~\eqref{scaleq1}
\begin{align}
\psi_1 &= \frac{1}{\sqrt{Z}} \prod_{j=1}^{N_\text{e}} \prod_{k=1}^{N} \left( \frac{\overline{z_j - \eta_k}\hphantom{^\alpha}}{|z_j - \eta_k|^\alpha}\right) \nonumber \\
&\times \prod_{i < j}^{N_\text{e}} (\bar{z}_i - \bar{z}_j) \exp\left(-\sum_{j=1}^{N_\text{e}} \frac{|z_j|^2}{4 l_B^2} \right),
\label{GSwf}
\end{align}
with the ground state energy $E_0 = N_\text{e} \hbar \omega_B/2$, independent of $\alpha$. For Eq.~\eqref{scaleq2}, the result is similar, with the substitution $\alpha \to \beta$. Here, $Z$ is the normalization constant, implicitly depending on the positions of the solenoids. It should be noted, for later convenience, that $\psi_{1,2}$ are single valued functions of both the variables $z_j$, as well as the parameters $\eta_k$.

\section{Geometric phases in the system}

In the previous section, we have found the ground state for the Hamiltonian $H = H_0 + V$, which is twofold degenerate. Now we study the evolution of GS as the solenoids are adiabatically moved around in the plane. Therefore, we study the time-dependent SE $H(t) |\Psi(t)\rangle = i\hbar |\dot{\Psi}(t)\rangle$, limited to the GS subspace. The Hamiltonian $H(t)$ becomes time-dependent due to the time dependence of the positions of the solenoids $\eta_k \to \eta_k(t)$. At any moment $t$, we can diagonalize the Hamiltonian according to $H(t) |\psi_n(t)\rangle = E_0 |\psi_n(t)\rangle$, with $|\psi_1(t)\rangle = \psi_1(t) \left|\uparrow\right\rangle$ and $|\psi_2(t)\rangle = \psi_2(t) \left|\downarrow\right\rangle$. By construction, this basis is orthonormal. Note that the energy eigenvalue $E_0$ does not have any time dependence because it does not depend on the positions of the solenoids. Now, let us track the evolution of two orthogonal states $|\Psi_n\rangle$, $n=1,2$, that were, initially, a specific linear combination of the energy basis eigenstates, $|\Psi_n(0)\rangle = C_{nm} |\psi_m(0)\rangle$ for some unitary matrix $C_{nm}$ (summation over repeated indices is implied).
By hypothesis, the evolution is adiabatic so that, at any time, we can write $|\Psi_n(t)\rangle \approx U_{nm}(t) |\psi_m(t)\rangle$. Substituting this ansatz into the time dependent SE, we find that the unitary operator $U_{nm}(t)$ has to satisfy the differential equation
\begin{equation}
\dot{U}_{nk} = U_{nm} \left(-i \frac{E}{\hbar} \delta_{mk} - i A_{mk} \right),
\label{DEUnit}
\end{equation}
with the initial condition $U_{nm}(0) = C_{nm}$. Here we have defined the Hermitian non-Abelian Berry connection as $A_{mk} = -i \langle \psi_k |\dot{\psi}_m \rangle$. The solution of the differential equation \eqref{DEUnit} is
\begin{equation}
U_{nk}(t) = \exp \left(-i \frac{E_0 t}{\hbar} \right) C_{nm}\, \mathcal{P}\! \exp \left(-i \int_0^t A_{mk}(t') d t' \right),
\label{UnitSol}
\end{equation}
where $\mathcal{P}\! \exp$ is a path ordered exponential operator. Now we are in the position to calculate the overlap of the evolved state with the initial state
\begin{equation}
\Gamma_{nm}(t) = \langle \Psi_m(0)| \Psi_n(t) \rangle = \bar{C}_{mr}U_{np}(t) \langle \psi_r (0)| \psi_p (t) \rangle.
\label{overlap}
\end{equation}
In general, $\langle \psi_r (0)| \psi_p (t) \rangle$ is not a unitary matrix, and neither is $\Gamma_{nm}(t)$. However, if we consider cyclic evolution of parameters $\eta_k$, so that after some time $t = \tau$, the Hamiltonian returns to its initial form, $H(\tau) = H(0)$, and so the eigenstates $\{|\psi_p(\tau)\rangle\}$ span the same subspace as $\{|\psi_r(0)\rangle\}$. Therefore, we may introduce the unitary matrix $\Delta_{pr} = \langle \psi_r (0)| \psi_p (\tau) \rangle$ which measures the degree of rotation of the eigenbasis after a complete cycle. With this, we finally arrive at the Wilczek-Zee matrix
\begin{equation}
\Gamma_{nm} \equiv \Gamma_{nm}(\tau) = U_{np}(\tau) \Delta_{pr}\bar{C}_{mr},
\label{WZmatrix}
\end{equation}
which contains all the information about the geometrical phases encoded in the system.

For the system at hand, the Wilczek-Zee matrix is greatly simplified in comparison to its most general form. First of all, the basis eigenstates $\{|\psi_n(t)\rangle\}$ are, by construction, single valued function of the positions of the solenoids $\eta_k$, so that a cyclic evolution $\eta_k(\tau) = \eta_k(0)$ implies $|\psi_n(\tau)\rangle = |\psi_n(0)\rangle$, which means that the matrix $\Delta_{pr} = \delta_{pr}$ is a unit matrix. Even greater simplification occurs due to the fact that, due to carefully chosen spin basis, the connection is diagonal
\begin{equation}
A_{mk} = 
\begin{pmatrix}
-i\int d^{2 N_\text{e}} z\, \bar{\psi}_1 \dot{\psi}_1 & 0 \\
0 & -i\int d^{2 N_\text{e}} z\, \bar{\psi}_2 \dot{\psi}_2
\end{pmatrix}_{mk},
\label{BerryConDiag}
\end{equation}
with entries that correspond to the Abelian Berry connections for two different couplings $\alpha$ and $\beta$ to the solenoid vector potential. In other words, our connection is a double copy of the Abelian Berry connection that was studied in \cite{Lunic2020}. Because of the diagonal form of the connection, the evolution operator can be explicitly calculated as
\begin{align}
&U_{nk}(\tau) = \exp \left(-i \frac{E_0 \tau}{\hbar} \right) C_{nm} \times \nonumber \\
&\underbrace{\begin{pmatrix}
  \exp \left[ -\int d^{2 N_\text{e}} z \oint\bar{\psi}_1 d \psi_1 \right]& 0 \\
0 &   \exp \left[-\int d^{2 N_\text{e}} z  \oint\bar{\psi}_2 d\psi_2 \right]
\end{pmatrix}_{mk}}_{B_{mk}}.
\label{UnitExpl}
\end{align}
Therefore, the Wilczek-Zee matrix for our system is of the form
\begin{equation}
\Gamma = \exp \left(-i \frac{E_0 \tau}{\hbar} \right) C B C^\dagger,
\label{GeoMat}
\end{equation}
where the $B$ matrix is the diagonal matrix describing the cyclic adiabatic evolution and $C$ matrix contains the information about initial conditions. In what follows, we will omit the dynamical phase factor $\exp \left(-i E_0 \tau/\hbar \right)$ so that we are left with purely geometric phase.

Before moving on to the main results of this paper, let us comment on some details regarding the adiabatic assumption. For this assumption to hold, the solenoids should move around slow enough so that their motion does not introduce additional energy to the system. The next available energy state, after the ground state, is the first excited state of the reduced Schr\"{o}dinger equation $ H_0 | \psi \rangle = E_1 \psi \rangle $. Therefore, one can say that the adiabatic assumption holds as long as the kinetic energies of the solenoids are less than $E_1 - E_0$. To give a more precise answer, one should know the energy of the first excited state $E_1$. This is, however, unobtainable by the methods we use to solve the ground state case. On the other hand, if the adiabatic assumption does not hold, the evolution of the system is no longer exclusively geometric in character and the desired anyonic interpretation of its behavior is lost.

\section{Results and discussion}

Having calculated the Wilczek-Zee matrix for the system, let us now investigate the possibility of interpreting the solenoids as non-Abelian anyons. (This was successfully done for the Abelian case in Ref.~\cite{Lunic2020}.) At first, the idea seems promising since the geometric phase of a system is described by the nontrivial Wilczek-Zee matrix. First of all, we calculate the trace of an arbitrary Wilczek-Zee matrix, i.e.~the Wilson loop. Using the cyclic property of the trace, we find that the unitary matrix $C$ makes no contribution to the trace and therefore
\begin{equation}
\mathrm{Tr}\, \Gamma = \mathrm{Tr}\, B.
\label{eq:Trace}
\end{equation}
Earlier we have shown that $B$ is a diagonal $2 \times 2$ matrix with different phases as its elements. Therefore, we have $\left| \mathrm{Tr}\, \Gamma \right| < 2$. This is a necessary but not a sufficient condition for the presence of non-Abelian anyons in the system \cite{Goldman2009, Goldman2014}. A sufficient condition is the existence of two different loops in the parameter space, $\gamma_1$ and $\gamma_2$, which share the starting point and give rise, via adiabatic evolution, to Wilczek-Zee matrices $\Gamma_1$ and $\Gamma_2$, respectively, that do not commute, $[\Gamma_1, \Gamma_2] \neq 0$. However, the Wilczek-Zee matrices obtained earlier fail this condition. Even though $\Gamma_{1,2}$ are not diagonal themselves, they are unitarily equivalent to the diagonal matrices $B_{1,2}$, i.e.~$\Gamma_{1,2} = C B_{1,2} C^\dagger$. This makes the commutator trivial
\begin{equation}
[\Gamma_1, \Gamma_2] = C [B_1, B_2] C^\dagger = 0,
\label{Commies}
\end{equation}
because the diagonal matrices always commute, $[B_1, B_2] = 0$. Therefore, there are no non-Abelian excitations in the system under consideration. Rather, there are two different kinds of Abelian anyons, which mix and carry a non-Abelian flux. A few comments are in order.

First, the diagonal nature of matrix $B$ is a direct consequence of the diagonal matrix $\mathcal{M}$ in the Hamiltonian. This would imply that systems of the type described by the Hamiltonian in Eqs.~\eqref{spHam} and \eqref{nonint} with a constant $\mathcal{M}$ do not contain non-Abelian anyons. In order to have a nonvanishing commutator $[\Gamma_1, \Gamma_2]$, one needs to have a nondiagonal matrix $B$, which requires the matrix $\mathcal{M}$ to be both (a) nondiagonal and (b) position dependent so that at each point in space, the diagonalization is performed by a different matrix. This means that, in the class of systems we are investigating, there would have to be a coupling between spin and position for the non-Abelian anyons to appear.

Second, it would certainly be possible to obtain the nonvanishing commutator for the two $\Gamma$ matrices if we chose different initial conditions $C$ for two different paths. However, such a commutator would lack any physical interpretation.

\begin{figure}%
\centering
\subcaptionbox{The loop $\gamma_1$ does not swap solenoids.}
{\includegraphics{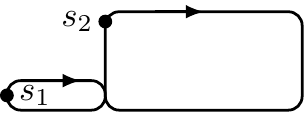}}
\hspace{15pt}
\subcaptionbox{The loop $\gamma_2$ swaps solenoids.}
{\includegraphics{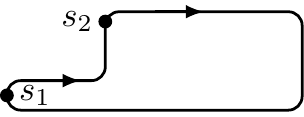}}
\caption{Two loops, $\gamma_1$ and $\gamma_2$, along which the solenoids $s_1$ and $s_2$ are adiabatically transported. Note that both loops span the same surface area so that no excess Aharonov-Bohm phase is introduced in the exchange phase \cite{Levin2003}.}%
\label{figura2}%
\end{figure}

Third, we can deduce the statistical phase obtained when we exchange the two solenoids, according to the procedure developed in Ref.~\cite{Levin2003}. To calculate this phase, we use the two different loops $\gamma_1$ and $\gamma_2$, shown in Fig.~\ref{figura2}, corresponding to unswapping and swapping of solenoids, respectively. If the corresponding Wilczek-Zee matrices are $\Gamma_1$ and $\Gamma_2$, then the statistical phase is contained in the matrix $S$ given by 
\begin{equation}
\Gamma_2 = S \Gamma_1.
\label{swapp}
\end{equation}
Due to the commuting nature of $\Gamma_1$ and $\Gamma_2$, and using the methods of Ref.~\cite{Lunic2020}, it can be verified that the matrix $S$ is simply given by
\begin{equation}
S = 
-\begin{pmatrix}
e^{i \pi \alpha} & 0 \\
0 & e^{i \pi \beta}
\end{pmatrix}.
\label{eq:matrixS}
\end{equation}
This result also confirms that the system carries non-Abelian flux and transforms simply, but nontrivially under the exchange of solenoids.

Finally, having discussed the theoretical predictions of our model, let us comment on its possible experimental implementation. Two most important features of our model can be realized in ultracold atomic gases~\cite{Bloch2008}: (1) the particles need to be confined in two spatial dimensions, and (2) they experience synthetic non-Abelian gauge potentials of external probes. The first one has been successfully implemented in numerous experiments (e.g., see \cite{Bloch2008} and references therein). As for the second feature, the first proposals for non-Abelian gauge potentials date more than fifteen years ago~\cite{Unanyan1999, Osterloh2005, Ruseckas2005}; however, non-Abelian gauge potentials have been successfully engineered only recently~\cite{Li2016, Bharath2019, DiLiberto2020, Sugawa2021}. Therefore, ultracold atomic gases seem like a promising platform for the experimental implementation of the Hamiltonian discussed here. For example, fine-tuned laser beams piercing the 2D ultracold atomic gas could be used in principle to create synthetic gauge potentials~\cite{Dalibard2011, Goldman2014, Lin2016}. However, at present, we are unable to propose a way to manipulate laser beams in order that they generate the specific form of the potential \eqref{nonAbPot}. Another viable route towards realization of non-Abelian vector potential in our model could be in specifically designed ion traps, as already presented recently in Ref.~\cite{Zygelman2021}: there, a toroidal trap simulates the motion of a planar rotor for a charged spin-$\frac{1}{2}$ ion, and an additional current and background magnetic field are present. In closing, we stress that, in this paper, we studied a model of synthetic anyons, rather than a concrete physical system with such properties.

\section{Conclusion}

In conclusion, we have presented an exactly solvable model of synthetic anyons in a 
many-body quantum system by employing external perturbations in the form solenoids carrying a non-Abelian flux. 
The model is fine-tuned so that each spin component of the electrons couples differently to the solenoids. 
To keep the model exactly solvable, we chose the simplest form of the non-Abelian potential of the solenoids that gives a non-Abelian flux and found the behavior to be similar to a double copy of the corresponding Abelian model. Although this model does not show the characteristic non-Abelian anyonic behavior under the adiabatic exchange of solenoids, it is, nevertheless, a convenient stepping stone in reaching such a model. Furthermore, similar Abelian systems with non-Abelian flux have recently sparked interest in their own right \cite{Alex2019}. It would be interesting to see how much information on the fully non-Abelian system can be extracted using this type of intermediate systems which are in between Abelian and non-Abelian anyons.

\section{Acknowledgments}

We acknowledge useful discussions with Robert Pezer. This work was supported by the QuantiXLie Center of Excellence, a project co-financed by the Croatian Government and European Union through the European Regional Development Fund - the Competitiveness and Cohesion Operational Programme (Grant KK.01.1.1.01.0004).



\begin{thebibliography}{99} 

\bibitem{Wilczek1982}
F. Wilczek,
Phys. Rev. Lett. {\bf 49}, 957 (1982).

\bibitem{Laidlaw1971}
M. G. G. Laidlaw and C. M. DeWitt,
Phys. Rev. D {\bf 3}, 1357 (1971).

\bibitem{Leinaas1977}
J. M. Leinaas and J. Myrheim,
Il Nuovo Cimento {\bf 37B}, 1 (1977).

%

\bibitem{Berry1984}
M. V. Berry,
Proc. R. Soc. Lond. A {\bf 392}, 45 (1984).

\bibitem{Nayak2008}	   
C. Nayak, S. H. Simon, A. Stern, M. Freedman, and S. Das Sarma,
Rev. Mod. Phys. {\bf 80}, 1083 (2008).

\bibitem{WZ1984}
F. Wilczek and A. Zee,
Phys. Rev. Lett. {\bf 52}, 2111 (1984).

\bibitem{Tsui1982}
D. C. Tsui, H. L. Stormer, and A. C. Gossard,
Phys. Rev. Lett. {\bf 48}, 1559 (1982).

\bibitem{Laughlin1983}
R. B. Laughlin,
Phys. Rev. Lett. {\bf 50}, 1395 (1983).

\bibitem{Arovas1984}
D. Arovas, J. R. Schrieffer, and F. Wilczek,
Phys. Rev. Lett. {\bf 53}, 722 (1984).

\bibitem{Camino2005}
F. E. Camino, Wei Zhou, and V. J. Goldman, 
Phys. Rev. B {\bf 72}, 075342 (2005).

\bibitem{Feldman2021}
D. E. Feldman, and B. I. Halperin,
Rep. Prog. Phys. {\bf 84} 076501 (2021).

\bibitem{Kitaev2003}
A. Y. Kitaev,
Ann. Phys. {\bf 303}, 2 (2003).

\bibitem{Kitaev2006}
A. Y. Kitaev,
Ann. Phys. {\bf 321}, 2 (2006).

\bibitem{Dai2017}
H.-N. Dai, B. Yang, A. Reingruber, H. Sun, X.-F. Xu, Y.-A.
Chen, Z.-S. Yuan, and J.-W. Pan, Nat. Phys. {\bf 13}, 1195 (2017).

\bibitem{Klanjsek2018}   
N. Jan\v{s}a, A. Zorko, M. Gomil\v{s}ek, M. Pregelj, K. W. Kr\"{a}mer,
D. Biner, A. Biffin, Ch. R\"{u}egg, and M. Klanj\v{s}ek,
Nat. Phys. {\bf 14}, 786 (2018).

\bibitem{DasSarma2015}	   
S. Das Sarma, M. Freedman, and C. Nayak,
npj Quantum Information {\bf 1}, 15001 (2015).

\bibitem{Mourik2012} 
V. Mourik, K. Zuo, S.M. Frolov, S.R. Plissard, E.P.A.M. Bakkers, L.P. Kouwenhoven, 
Science {\bf 336}, 1003 (2012).

\bibitem{Paredes2001}
B. Paredes, P. Fedichev, J. I. Cirac, and P. Zoller,
Phys. Rev. Lett. {\bf 87}, 010402 (2001).

\bibitem{Zhang2014}
Y. Zhang, G. J. Sreejith, N. D. Gemelke, and J. K. Jain,
Phys. Rev. Lett. {\bf 113}, 160404 (2014).

\bibitem{Duan2003}
L.-M. Duan, E. Demler, and M. D. Lukin,
Phys. Rev. Lett. {\bf 91}, 090402 (2003).

\bibitem{Jiang2008}
L. Jiang, G. K. Brennen, A. V. Gorshkov, K. Hammerer, M. Hafezi, E. Demler, M. D. Lukin, and P. Zoller, 
Nat. Phys. {\bf 4}, 482 (2008).

\bibitem{Burrello2010}
M. Burrello, and A. Trombettoni,
Phys. Rev. Lett. {\bf 105}, 125304 (2010).

\bibitem{Andrade2021}
B. Andrade, V. Kasper, M. Lewenstein, C. Weitenberg, and T. Gra{\ss},
Phys. Rev. A {\bf 103}, 063325 (2021).

\bibitem{Baldelli2021}
N. Baldelli, B. Juli\'{a}-D\'{i}az, U. Bhattacharya, M. Lewenstein, and T. Gra{\ss},
Phys. Rev. B {\bf 104}, 035133 (2021).

\bibitem{Kapit2014}
E. Kapit, M. Hafezi, and S. H. Simon,
Phys. Rev. X {\bf 4}, 031039 (2014).

\bibitem{Umucalilar2017}
R. O. Umucalilar, and I. Carusotto, 
Phys. Rev. A {\bf 96}, 053808 (2017).

\bibitem{Lu2009}
C.-Y. Lu, W.-B. Gao, O. Guhne, X.-Q. Zhou, Z.-B. Chen, and J.-W. Pan, 
Phys. Rev. Lett. {\bf 102}, 030502 (2009).

\bibitem{Pachos2009}
J. K. Pachos, W. Wieczorek , C. Schmid, N. Kiesel, R. Pohlner, and H. Weinfurter, 
New J. Phys. {\bf 11}, 083010 (2009).

\bibitem{Zhong2016}
Y.-P. Zhong, D. Xu, P. Wang, C. Song, Q.-J. Guo, W.-X. Liu, K. Xu, B.-X. Xia, C.-Y. Lu, S. Han, J.-W. Pan, and H. Wang, 
Phys. Rev. Lett. {\bf 117}, 110501 (2016).

\bibitem{Li2017}
K. Li, Y. Wan, L.-Y. Hung, T. Lan, G. Long, D. Lu, B. Zeng, and R. Laflamme,
Phys. Rev. Lett. {\bf 118}, 080502 (2017).

\bibitem{Papic2018}
Z. Papi\'{c}, R. S. K. Mong, A. Yazdani, and M. P. Zaletel,
Phys. Rev. X {\bf 8}, 011037 (2018).

\bibitem{Jaeck2021}
B. J\"{a}ck, Y. Xie, and A. Yazdani,
Nat. Rev. Phys {\bf 3}, 541--554 (2021).

\bibitem{Brooks2021}
M. Brooks, M. Lemeshko, D. Lundholm, and E. Yakaboylu,
 Phys. Rev. Lett. {\bf 126}, 015301 (2021).

\bibitem{Nakamura2020}
J. Nakamura, S. Liang, G. C. Gardner, and M. J. Manfra,
Nat. Phys. {\bf 14}, 931--936 (2020).

\bibitem{Bartolomei2020}
H. Bartolomei, M. Kumar, R. Bisognin, A. Marguerite, J. M. Berroir, E. Bocquillon, B. Pla\c{c}ais, A. Cavanna, Q. Dong, U. Gennser, Y. Jin, and G. F\`{e}ve,
Science {\bf 368}, 173--177 (2020).

\bibitem{Noh2020}
J. Noh, T. Schuster, T. Iadecola, S. Huang, M. Wang, K. P. Chen, C. Chamon, and M. C. Rechtsman,
Nat. Phys. {\bf 16}, 989--993 (2020).

\bibitem{Barreiro2011}
J. T. Barreiro, M. M\"{u}ller, P. Schindler, D. Nigg, T. Monz, M. Chwalla, M. Hennrich, C. F. Roos, P. Zoller, and R. Blatt,
Nature {\bf 470}, 486 (2011).

\bibitem{Yang2019}
Y. Yang, C. Peng, D. Zhu, H. Buljan, J. D. Joannopoulos, B. Zhen, and M. Solja{\v c}i{\'c},
Science {\bf 365}, 6457 (2019).

\bibitem{Weeks2007}
C. Weeks, G. Rosenberg, B. Seradjeh, and M. Franz, 
Nat. Phys. {\bf 3}, 796 (2007).

\bibitem{Rosenberg2009}
G. Rosenberg, B. Seradjeh, C. Weeks, and M. Franz, 
Phys. Rev. B {\bf 79}, 205102 (2009).

\bibitem{Klitzing1980}
K. v. Klitzing, G. Dorda, and M. Pepper,
Phys. Rev. Lett. {\bf 45}, 494 (1980).

\bibitem{Laughlin1981}
R. B. Laughlin, 
Phys. Rev. B {\bf 23}, 5632 (1981).

\bibitem{Rahmani2013}
A. Rahmani, R. A. Muniz, and I. Martin, 
Phys. Rev. X {\bf 3}, 031008 (2013).

\bibitem{SeradjehPRL2008}
B. Seradjeh, and M. Franz, 
Phys. Rev. Lett. {\bf 101}, 146401 (2008).

\bibitem{Todoric2018}
M. Todori\'{c}, D. Juki\'{c}, D. Radi\'{c}, M. Solja\v{c}i\'{c}, and H. Buljan, 
Phys. Rev. Lett. {\bf 120}, 267201 (2018).

\bibitem{Lunic2020}
F. Luni\'{c}, M. Todori\'{c}, B. Klajn, T. Dub\v{c}ek, D. Juki\'{c}, and H. Buljan, 
Phys. Rev. B {\bf 101}, 115139 (2020).

\bibitem{Todoric2020}
M. Todori\'{c}, B. Klajn, D. Juki\'{c}, and H. Buljan,
Phys. Rev. A {\bf 102}, 013322 (2020).

\bibitem{Wu1975}
T. T. Wu, and C. N. Yang,
Phys. Rev. D. {\bf 12}, 3845 (1975).

\bibitem{Zygelman2021}
B. Zygelman,
Phys. Rev. A {\bf 103} 042212 (2021).

\bibitem{Goldman2009}
N. Goldman, A. Kubasiak, P. Gaspard, and M. Lewenstein, 
Phys. Rev. A {\bf 79}, 023624 (2009).

\bibitem{Goldman2014}
N. Goldman, G. Juzeli\={u}nas, P. \"{O}hberg, and I. B. Spielman, 
Rep. Prog. Phys. {\bf 77}, 126401 (2014).

\bibitem{Levin2003}
M. Levin, and X. G. Wen, 
Phys. Rev. B {\bf 67}, 245316 (2003).

\bibitem{Bloch2008}
I. Bloch, J. Dalibard, and W. Zwerger,
Rev. Mod. Phys. {\bf 80}, 885 (2008).

\bibitem{Unanyan1999}
R. G. Unanyan, B. W. Shore, and K. Bergmann,
Phys. Rev. A {\bf 59}, 2910 (1999).

\bibitem{Osterloh2005}
K. Osterloh, M. Baig, L. Santos, P. Zoller, and M. Lewenstein,
Phys. Rev. Lett. {\bf 95}, 010403 (2005).

\bibitem{Ruseckas2005}
J. Ruseckas, G Juzeliunas, P. \"{O}hberg, and M. Fleischhauer,
Phys. Rev. Lett. {\bf 95}, 010404 (2005).

\bibitem{Li2016}
T. Li, L. Duca, M. Reitter, F. Grusdt, E. Demler, M. Endres, M. Schleier-Smith, I. Bloch, and U. Schneider,
Science {\bf 352}, 6289, 1094--1097 (2016).

\bibitem{Bharath2019}
H. M. Bharath, M. Boguslawski, M. Barrios, L. Xin, and M. S. Chapman,
Phys. Rev. Lett. {\bf 123}, 173202 (2019).

\bibitem{DiLiberto2020}
M. Di Liberto, N. Goldman, and G. Palumbo,
Nat. Comm. {\bf 11}, 5942 (2020).

\bibitem{Sugawa2021}
S. Sugawa, F. Salces-Carcoba, Y. Yue, A. Putra, and I. B. Spielman,
npj Quantum Information {\bf 7}, 144 (2021).

\bibitem{Dalibard2011}
J. Dalibard, F. Gerbier, G. Juzeliunas, and P. \"{O}hberg, 
Rev. Mod. Phys. {\bf 83}, 1523 (2011).

\bibitem{Lin2016}
Y. J. Lin, and I. B. Spielman,
J. Phys. B {\bf 49}, 183001 (2016).

\bibitem{Alex2019}
M. Kremer, L. Teuber, A. Szameit, and S. Scheel, 
Phys. Rev. Res {\bf 1}, 033117 (2019).

%
%
%
%
%
%

%

%
%
%
%
%


%



\end{thebibliography}
\end{document}